%% file: main.tex
\documentclass[manuscript]{acmart}

\setcopyright{none}
\usepackage{booktabs}
\usepackage{graphicx}
\usepackage{amsmath}
\usepackage{algorithm}
\usepackage{algpseudocode}
\usepackage{wrapfig}
\usepackage{multirow}

\title{Application-Defined Abstractions to Separate Numerical Intent from Computation Orchestration} 

\author{Elena Aschmann}
\affiliation{%
\institution{Swiss Federal Institute of Technology}
\city{Zurich}
\country{Switzerland}
}
\email{easchmann@ethz.ch}
\author{Anshu Dubey}
\affiliation{%
\institution{RIKEN Center for Computational Science}
\city{Kobe}
\country{Japan}
}
\email{anshu.dubey@riken.jp}
\author{Mohamed Wahib}
\affiliation{%
\institution{RIKEN Center for Computational Science}
\city{Tokyo}
\country{Japan}
}
\email{mohamed.attia@riken.jp}
\author{Jens Domke}
\affiliation{%
\institution{RIKEN Center for Computational Science}
\city{Kobe}
\country{Japan}
}
\email{jens.domke@riken.jp}

\begin{document}

\begin{abstract}
In High-performance Computing the holy grail has been to translate numerical algorithms expressed in mathematical notation into code that performs well across a range of platforms. A variety of  portability frameworks have abstracted data layout, parallelization and orchestration  from application's view, however, the resulting format is still code that conforms to the idioms of the framework. Additionally, substantial refactoring is typically needed to achieve performance goals. Increasing sophistication of Large Language Models (LLMs) naturally raises the question whether they can be used to make application codes and/or solvers both portable and performant. The obstacle is the lack of correctness guarantee in LLM produced code. In this paper we pose the question differently -- can we start with expressing numerical intent of the code into an immutable core arithmetic component isolated from the surrounding code such that the surrounding code can be safely generated by the LLMs? We answer this question  by using Spark, a mature compressible hydrodynamics solver in Flash-X, a multiphysics software system. We use LLM assistance in devising the abstractions and embedding them in the solver. Once properly verified, the resulting solver is lowered to concrete realization deterministically with tools also built with LLM assistance. The driver of the numerics and the code for interfacing with the application are generated by the LLMs. The resulting solver is exercised within Flash-X and with a separate environment we built for this experiment. Our results show that this is a viable approach for producing correct and performance portable code. 
\end{abstract}

%


\begin{teaserfigure}
    \begin{center}
   \includegraphics[width=\linewidth]{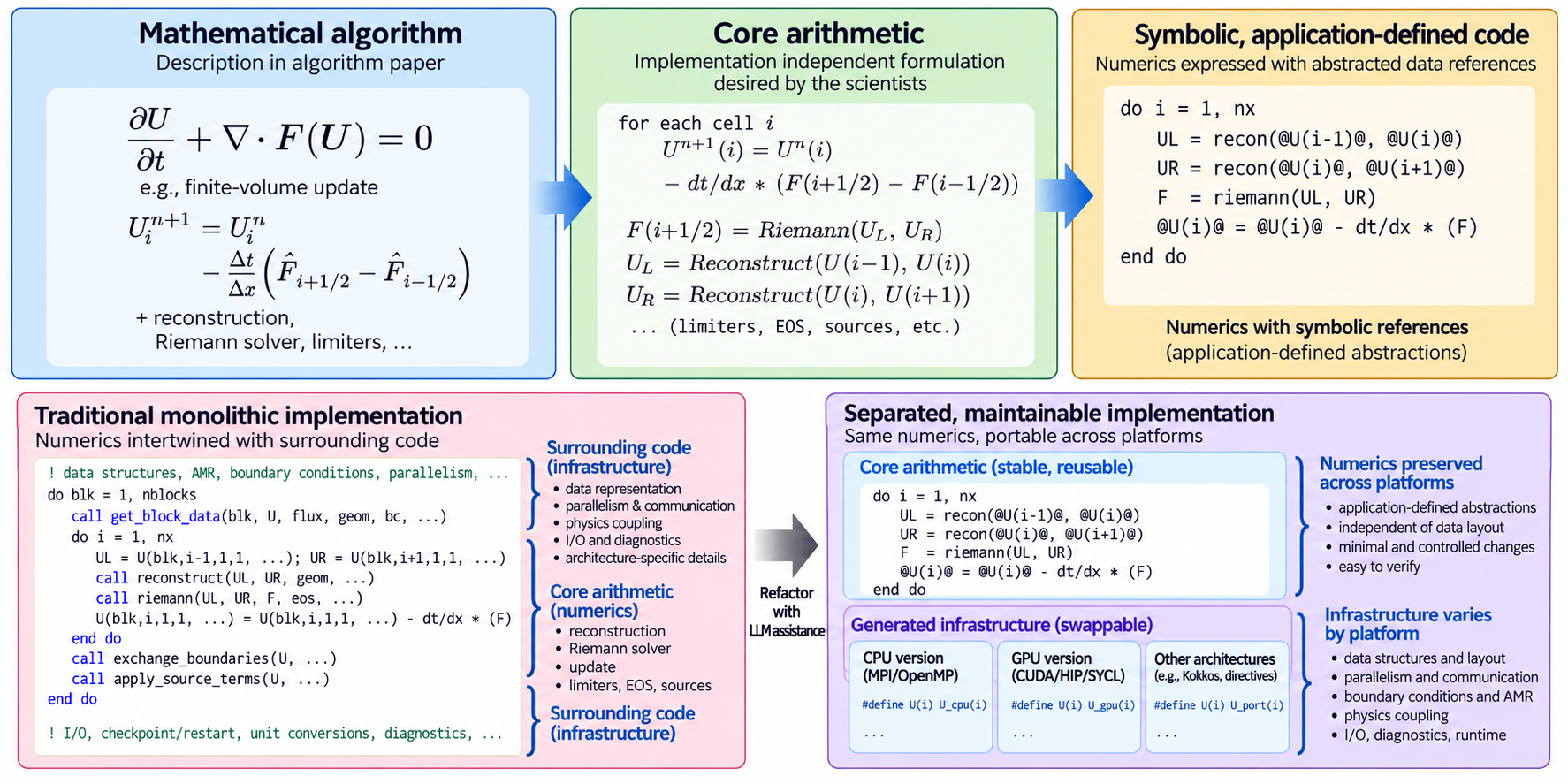}
    \caption{ A methodology to achieve separation of the numerical intent of the solver from its surrounding code through application defined and controlled abstractions}
    \label{fig:teaser}
\Description{Diagram showing the separation of core arithmetic
  from infrastructure through application-defined abstractions.}     \end{center}
\end{teaserfigure}

\maketitle

\section{Introduction}
\label{sec:intro}
For domain scientists and applied mathematicians working in high-performance computing (HPC) environments the holy grail has been to express their algorithm in purely mathematical notation that can be used to generate code that performs well across a range of platforms. The increasing sophistication of large language models (LLMs), and agents that drive them, have brought us close to achieving this goal. Figure \ref{fig:teaser} summarizes the premise and the process involved. Blue box shows how an algorithm might be presented in mathematical notation in a paper. The green box describes step-by-step arithmetic needed to implement the algorithm -- a representation preferred by the domain scientists and computational mathematicians. This description is implementation independent. The yellow box shows how the representation in green box can be converted to a code with embedded abstractions that are solver or application specific. These abstractions, and the tools for lowering them to target language of the code are developed with LLM assistance. The pink box shows what corresponding traditional code might have looked like, highlighting portions that are core arithmetic versus surrounding code. The lavender box shows the core arithmetic that is translated using a verified tool and the variants of the surrounding code that are generated using LLMs. Our ultimate goal is to go from blue box to lavendar box, but for this work we begin from the green box. We start with a mature solver, extract its arithmetic into mathematical notation and use that as the basis for generating a ``new'' implementation. We take this approach because with a well tested solver we have a way to validate our hypothesis. 

We use Spark \cite{Couch2021}, a solver for compressible hydrodynamics, embedded in a large multiphysics software system, Flash-X \cite{DubeyEtAl22} for the experiment.  We first extract the description of the algorithm in a detailed mathematical notation similar to what is shown in the green box in figure \ref{fig:teaser}. Then we convert this representation of the core arithmetic into a succinct and reusable form that is independent of the surrounding infrastructure necessary to facilitate the computation. Components of such infrastructure include data representation and movement, coordinating interactions among physics components, exposing parallelism, managing communication, and accommodating the constraints of different computing architectures. We refer to this infrastructure as {\em surrounding code} and distilled numerical component of the code as {\em core arithmetic} from here on for ease of explanation.  With this exercise we wished to  explore how far could we push code generation with LLMs to produce reliable code suitable for different platforms. Our finding is that it is possible to design, build and maintain application-specific abstractions that distance the core arithmetic from surrounding code up to the degree of intrusiveness acceptable to the developers. 

Such a separation has long been desirable for maintainability and portability, and several efforts over the years have achieved it to varying degrees \cite{Dai2024PerformancePortability,Edwards2014,Ren2026Kokkos,GridTools,legion,clement2018claw,gysi2015stella,halide}. Fully realizing this separation has not been possible because application-specific abstractions were not viable as long as every line of code had to be written by humans. The breath and depth of expertise necessary to design, develop, and maintain abstractions is rare enough that groups engaged in these activities had to necessarily target a broad range of applications. The non-specificity resulting from this constraint led naturally to libraries that are large and complex and require significant resources to maintain and keep up to date with evolution in technology.  

From the applications perspective also these solutions only provide moderate benefits. They have their own type system, data structures, and programming idioms which the application must adopt, and even then obtaining performance requires the application structure to be suitable for the target architecture. AI-assisted code development \cite{Fan2023,Hou2024,Jiang2026} could change this trade-off. However, even a fully validated specification cannot simply be handed over to an LLM to produce concrete code directly because of the fundamental limitations in the abilities of the models to guarantee correct outcome. The key, therefore, is to convert the mathematical notation to application specified abstractions and use LLMs to build tools that can lower the abstractions to a concrete realization in the target language. Once the abstractions and tools have been validated the translation is deterministic and by definition guaranteed to be correct up to the rigor of the verification applied to the translation. Because the abstraction boundary is controlled by the application rather than by an external framework, it can evolve with the application and can be introduced incrementally into an existing code base.

In this paper, we develop and demonstrate this approach with Spark and use two different sets of abstractions to demonstrate the range and flexibility of the approach. Computational kernels are expressed using application-defined abstractions in both versions, though at different levels of intrusiveness into the original code. 

The contributions of this work are:
\vspace{-0.1in}
\begin{itemize}
    \item Defining abstraction boundary that guarantees reliable outcome with LLM assisted code generation.
    \item A stable, infrastructure-agnostic core arithmetic component where explicit data references can be replaced by symbolic, parameterized references, creating a maintainable scientific artifact.
    \item AI-assisted transformation of existing scientific software in a  workflow where use of generative techniques is shifted from repeatedly generating scientifically sensitive code to performing constrained, verifiable source transformations.
    \item Demonstration of correctness across independently developed infrastructures and performance on different architectures.
\end{itemize}

\section{Premise}
Large scientific simulation codes differ in important ways from conventional application software in that scientific knowledge is encoded in the implementations whose correctness has been established over many years of scientific use. This is the primary reason why LLM assisted development has not made as many inroads in scientific computing as it has elsewhere. However, the scientifically meaningful core arithmetic of the solvers on which the reliability of the obtained results hinges is typically a small fraction of the overall code. Therefore, if this core could be effectively isolated and made immutable then LLMs can be used for generating the surrounding code. Given the lack of guaranteed correctness of LLM generated code, any adoption of generative practices must be accompanied by rigorous verification. 

The challenge has been that the organization of data is usually embedded in numerical source code through explicit array subscripts and derived-type component selections. Consequently, a change in the data layout requires changes throughout the code, therefore, any LLM generated surrounding code would also entangle the arithmetic. The premise behind our experiment is that completely detaching the core arithmetic from the surrounding code is possible and the exercise of detachment will result in a precise elucidation of the scientific core of the code without being contaminated by details of the infrastructure that frames it. This can be done by defining a vocabulary of symbols that have specific meanings within the context of the application. Symbols can be as limited as data references only, or they can represent arbitrarily sized code-snippets with parameterization. One then build tools to lower these symbols into the concrete source-level expression required by a particular environment. Once verified and validated, this representation becomes the enduring and stable scientific representation of the application. 

The resulting development model differs from both conventional abstraction libraries and unconstrained AI code generation. The scientifically significant arithmetic remains an explicit, maintained software artifact. The application itself determines the abstraction, AI assists in creating and manipulating this representation, while deterministic code generation lowers the abstractions into infrastructure-specific source code. This division provides the basis for the approach demonstrated in the remainder of this paper.

\section{Instruments of Experimentation}
We give a brief outline of the solver and the code used in this study to facilitate the description of the methodology. Flash-X is a component-based software system for building multiphysics simulation applications where different permutations of code components generate specific instantiations capable of simulating the requisite physics. Components can have multiple alternative implementations to account for either different physical regimes or different fidelity requirements. It can be run with a uniformly discretized mesh (UG) or with adaptive mesh refinement (AMR) where higher resolution is used in the mesh only where necessary. Flash-X has a  customized macro-expander \cite{Dubey23} that became the starting point for the abstractions derived in this work. These custom macros can have multiple alternative definitions which allow them to unify code variants into one source. The expander has a built-in ability to arbitrate on which definition to use where to meet the requirements of various configurations. 

Spark solves the equations of ideal MHD using finite-volume
discretization and explicit time integration, although in this work we disable the magnetic field component of the solver and use only the hydrodynamics. The solver begins by performing high-order spatial reconstruction on the cell-averaged primitive variables to estimate their values at the interface between cells. Running offset stencils for this reconstruction results in two approximations for the vector of primitives at each interface. These left and right states at each interface are used as input to a Riemann solver to compute the average flux across each interface between cells.  Then the interface fluxes are used to compute conservative flux divergences for the update of the field variables. Spark uses second- or third-order strong stability preserving (SSP) Runge-Kutta (RK) time integration \cite{shu:1988,gottlieb:2009}.  This approach requires multiple sub-stages in which the vector of conserved quantities is updated requiring that the general procedure outlined above be repeated two or three times depending on whether second- or third-order time integration is used.  The default for reconstruction is fifth-order weighted essentially non-oscillatory (WENO) \cite{shu:2009} with the smoothness indicators of \cite{borges:2008}, and the HLLD approximate Riemann solver \cite{miyoshi:2005}. These are the calculations that form the core arithmetic of the solver that we extract and to which we apply the abstractions.

\begin{figure}[htbp]
\vspace{-0.2in}
    \centering
    \includegraphics[width=0.95\textwidth]{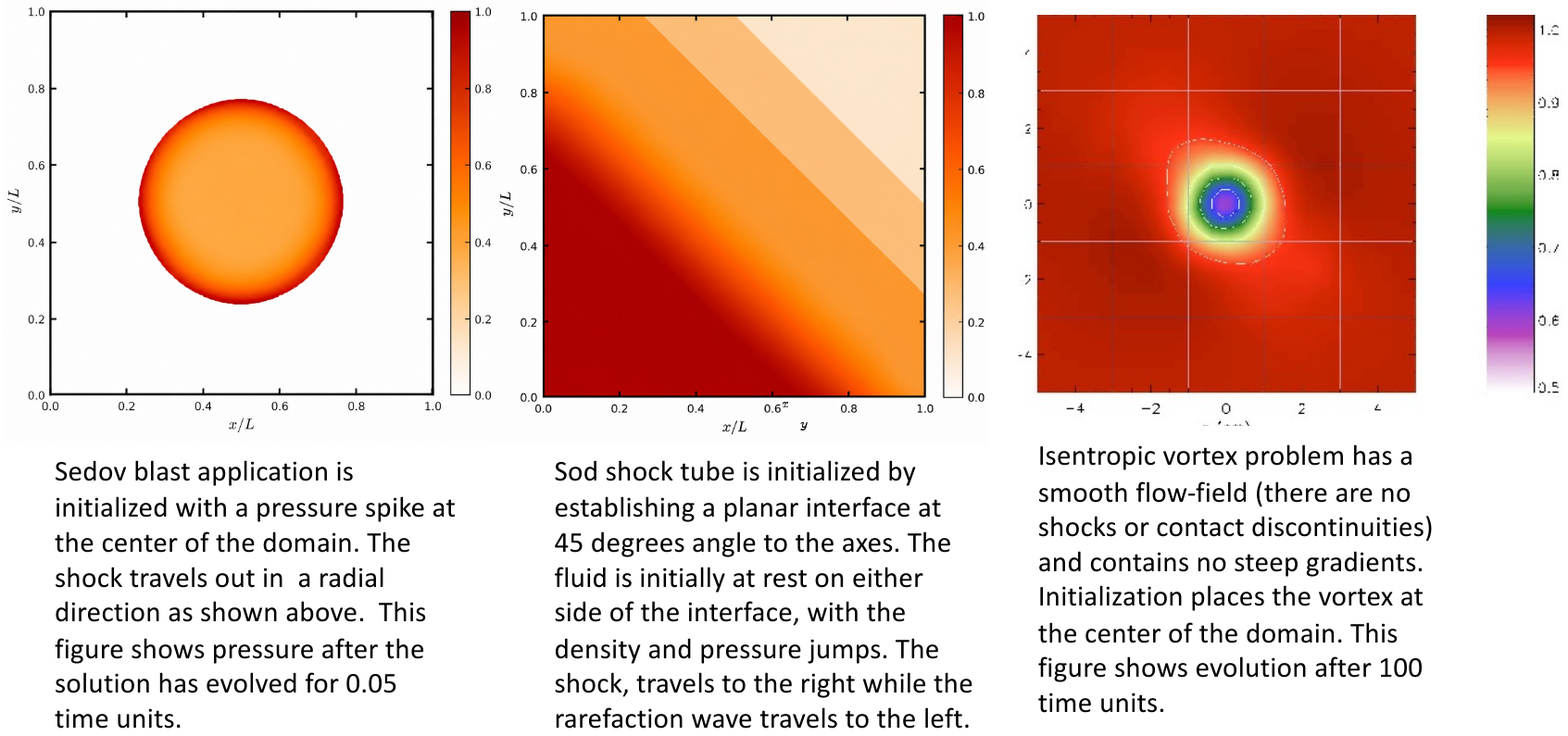}
    \captionsetup{skip=-35pt}    
    \caption{General visualizations of Sedov blast, Sod-shocktube and Isentropic-vortex problems after they have evolved for a while}
    \label{fig:problems}
\end{figure}

We use three different applications in this study that are considered standard problems for exercising hydrodynamics solvers. They are, \texttt{Sedov blast} \cite{sedov1993}, \texttt{Sod} shock tube \cite{Sod1978} and \texttt{IsentropicVortex} \cite{Shu1997} shown in Figure \ref{fig:problems}. \footnote{We reproduce the visualizations from Flash-X User's Guide to give the reader an idea of what these problems are like, these are not produced from our experiments}. The Sedov blast problem initializes the domain with a pressure spike where the shock wave moves out radially. Sod produces all three types of nonlinear  hydrodynamic waves (shock, contact, and rarefaction) associated with a shock, and isentropic vortex has a smooth profile with no shocks. All three applications use ideal gas equation of state.  

We used both commercially hosted and local institution hosted models for this work. The commercially hosted model were GPT-6.1-sol light with Codex as the agent and Opus-5.5 Medium. Institutionally hosted models included DeepSeek-V4 and GLM-5.3. Our objective in this exercise is not to focus on any one model. We list all the models that we used for transparency and completeness.

\input methodology

\input experiments

\input results
\section{Related Work}

This work lies at the boundary of performance-portability abstractions for HPC applications where code-generation approaches separate numerical intent from implementation, and AI-assisted transformation of existing software.  Extant performance portability approaches have concerned themselves with providing abstractions for parallel execution, memory spaces, and data access while mapping these abstractions onto architecture-specific backends \cite{Trott2021Kokkos,Zenker2016,Stephan2026}. Directive-based approaches such as OpenMP \cite{openmp} and OpenACC \cite{openacc} target offloading allow architecture mappings to be expressed through annotations rather than by adopting a C++ abstraction library. The commonality among them is that the portability mechanism defines the abstractions available to the application. 

LLMs have introduced a new mechanism for automating software transformations that previously required substantial manual effort. However, the reliability of such transformations remains a central concern \cite{Martinez2025, LLMCodeQuality2026}. This is particularly important for scientific software, where apparently small changes to arithmetic expressions, array indexing, or update ordering can silently alter numerical results. 

Our study shows that a constrained use of LLMs with a robust verification framework can be used to shift the abstraction boundary to the application. The application defines a small symbolic vocabulary corresponding to its own existing computational and data-access constraints. Infrastructure-specific code generation then gives those symbols their concrete representation. Consequently, an application's numerical implementation need not adopt the types, containers, or access syntax of a particular portability framework. It also differs from conventional source-to-source porting. Rather than transforming an application directly from one infrastructure-specific representation to another, there is an intermediate, application-controlled representation into the maintained source. Once this transformation has been performed and validated, subsequent infrastructure changes alter the lowering of the abstractions rather than the arithmetic that expresses the numerical method.

\section{Conclusion}
\label{sec:conclusion}
We have introduced a new way of thinking about abstractions where instead of conforming to the a general portability system's requirements, the applications devise their own customized portability framework. Until the arrival of highly capable LLMs and agentic frameworks this wasn't feasible. Each transformation had to be manually constructed and validated. The other extreme of using LLM to produce all the code is still not feasible because of reliability concerns. The central proposition of this work is not that scientific software should be regenerated whenever computing technology changes. Rather, the availability of reliable code-generation assistance allows us to reconsider where the stable boundary in scientific software should lie. By maintaining the scientifically meaningful arithmetic as a durable artifact and making its connection to infrastructure explicit through application-defined abstractions, it is possible to preserve validated scientific computation while allowing the surrounding software mechanisms to evolve with substantially less disruption. The GPU configuration tests this proposition directly. Moving the arithmetic to a GPU, and making it efficient there by batching blocks required only new surrounding code. The arithmetic layer stayed the same and reproduced the CPU results to rounding; with contraction disabled, the GPU and host executions of the generated kernels were bit-wise identical.

We have exercised this proposition by introducing two levels of abstractions into an existing, scientifically validated implementation. One is a shallow abstraction that only affects array references leaving the remainder of the implementation unaltered, while the other is a fundamentally different way of implementing an algorithm from its detailed mathematical exposition. In the second approach the role of the reference implementation is limited. It is used to extract the detailed mathematical specification of the algorithm and as a baseline for verifying the generated code.  Our ultimate objective is to use LLM assistance to go from the detailed mathematical description of the algorithm to generated executable code. Here we use extraction of mathematical notation from a mature solver because having a reference implementation allows us to validate the methodology. It is also worth noting that with this approach the degree of intrusiveness of the abstractions remains firmly in control of the application developer.


%


\bibliographystyle{ACM-Reference-Format}
\bibliography{references}

\end{document}

%% file: methodology.tex
\section{Abstracting Data Layout}
\label{sec:dataref-only}
The most obvious abstraction, and the one embraced by various C++ based performance portability solutions such as Kokkos \cite{Edwards2014,Trott2021Kokkos,Ren2026Kokkos}, is to eliminate explicit data references from the arithmetic. We use a macro-like mechanism explained in \cite{Dubey23} for these simplest abstractions . Similar to macros, these symbols are strings with parameters that are replaced with their pre-defined expansions by the translator.
For example, the code below can be translated into several different versions depending on how the symbols \texttt{spc}, \texttt{spc\_off}, \texttt{field}, and \texttt{field\_off} are defined.
\begin{verbatim}
rho(@spc(i,j,k)@) = U(@field(DENS,i,j,k)@)
flux(@spc(i,j,k)@) = &
    F(@field_off(DENS,i,+1,j,+0,k,+0)@) - &
    F(@field_off(DENS,i,-1,j,+0,k,+0)@)
\end{verbatim}
Table~\ref{table:tran} lists symbol definitions and their corresponding expansions into Fortran code. The first three columns show the symbol, its arguments and how it is referenced in the code snippet. The last two columns show how changing the definition changes the code translation and what goes into the source code. The juxtaposed arguments are concatenated so spc\_off(i,+1,j,+0,k,+0) yields i+1,j+0, k+0. These examples illustrate that the symbolic representation preserves the logical identity and relative offsets of data references, while the mapping from logical indices to their physical ordering is determined entirely by how the symbols are defined. The entire transformation can be verified by asking for a set of definitions that are compatible with the original data layout of the code. Lowered to the level of assembly, the original code and the translated code with embedded symbols match exactly. Formal methods can also be applied to verify the semantic equivalence of any alternative set of definitions representing a different data layout \cite{siegel2015civl}, though we did not use them here because the variants are simple enough to produce bit-wise identical outputs. 

\begin{table}[htbp]
\centering
\caption{Translation of application-defined symbols using two
alternative definitions. For each symbol, the first row shows
the original spatial index ordering and the second row shows
the reversed ordering.}
\label{table:tran}
\scriptsize
\resizebox{\textwidth}{!}{%
\begin{tabular}{|l|l|l|l|l|}
\hline
Symbol & Arguments & Reference & Definition & Translation \\
\hline

\multirow{2}{*}{\texttt{spc}} &
\multirow{2}{*}{\texttt{ib,jb,kb}} &
\multirow{2}{*}{\texttt{\detokenize{@spc(i,j,k)@}}} &
\texttt{ib,jb,kb} &
\texttt{i,j,k} \\
\cline{4-5}
& & &
\texttt{kb,jb,ib} &
\texttt{k,j,i} \\
\hline

\multirow{2}{*}{\texttt{spc\_off}} &
\multirow{2}{*}{\texttt{ib,io,jb,jo,kb,ko}} &
\multirow{2}{*}{\texttt{\detokenize{@spc_off(i,+1,j,-1,k,+0)@}}} &
\texttt{ibio,jbjo,kbko} &
\texttt{i+1,j-1,k+0} \\
\cline{4-5}
& & &
\texttt{kbko,jbjo,ibio} &
\texttt{k+0,j-1,i+1} \\
\hline

\multirow{2}{*}{\texttt{field}} &
\multirow{2}{*}{\texttt{var,ib,jb,kb}} &
\multirow{2}{*}{\texttt{\detokenize{@field(DENS,i,j,k)@}}} &
\texttt{\detokenize{var,@spc(ib,jb,kb)@}} &
\texttt{DENS,i,j,k} \\
\cline{4-5}
& & &
\texttt{\detokenize{var,@spc(ib,jb,kb)@}} &
\texttt{DENS,k,j,i} \\
\hline

\multirow{2}{*}{\texttt{field\_off}} &
\multirow{2}{*}{\texttt{var,ib,io,jb,jo,kb,ko}} &
\multirow{2}{*}{\texttt{\detokenize{@field_off(DENS,i,-1,j,+0,k,+0)@}}} &
\texttt{\detokenize{var,@spc_off(ib,io,jb,jo,kb,ko)@}} &
\texttt{DENS,i-1,j+0,k+0} \\
\cline{4-5}
& & &
\texttt{\detokenize{var,@spc_off(ib,io,jb,jo,kb,ko)@}} &
\texttt{DENS,k+0,j+0,i-1} \\
\hline

\end{tabular}%
}
\end{table}


To apply this methodology to Flash-X we used the already existing macro expansion mechanism reported in \cite{Dubey23}. 
We needed an alternative implementation of the mesh to test the mutability of the data layout handled in the solver. With AI assistance we generated an alternative implementation of Uniform Grid in the code which changes the order of indices from variable first to variable last (see section \ref{sec:verification} for more details). We verified correctness by comparing the output against that produced by the unchanged solver running on the unchanged Uniform Grid. 

\section{Implementation from Mathematical Notation}
In the next stage of exploration we pushed the abstractions deeper into the code. Here, we completely detach core arithmetic from the surrounding code. 

\begin{wrapfigure}{r}{0.46\linewidth}
    \centering
    \vspace{-10pt}
    \includegraphics[
        width=\linewidth,
        trim=20 100 280 100,
        clip
    ]{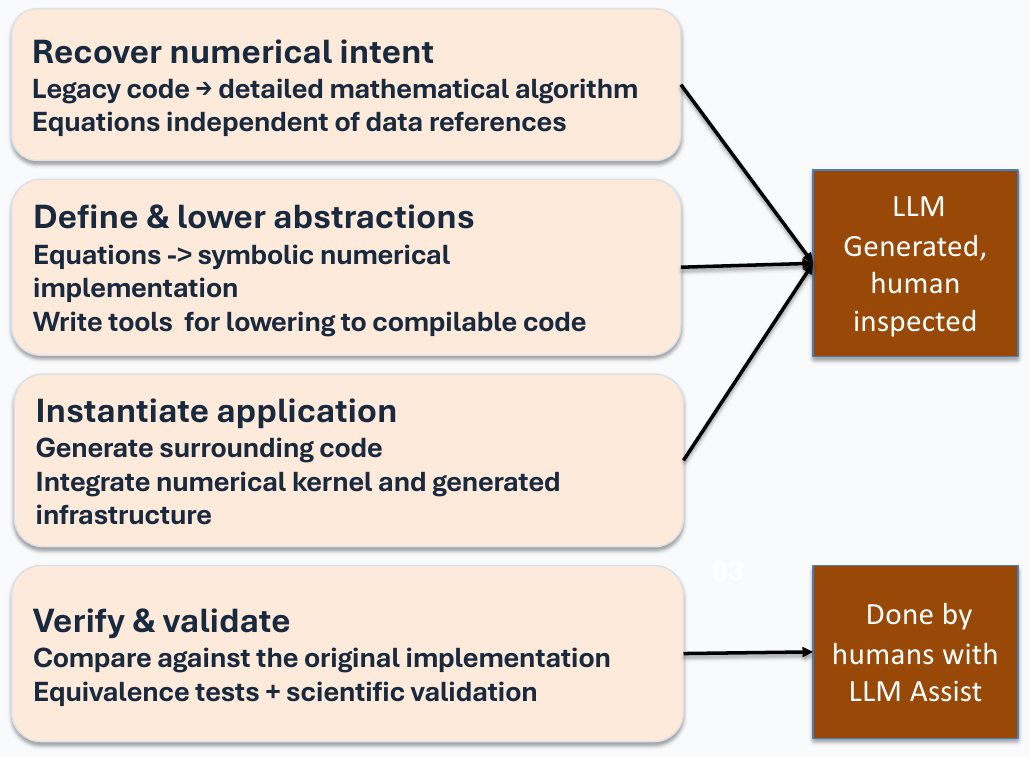}
    \caption{Steps involved in algorithmic decomposition for
    separation of concerns, including the roles apportioned
    to AI and human scientists.}
    \label{fig:schem}
    \vspace{-10pt}
\end{wrapfigure}

Figure~\ref{fig:schem} summarizes the steps in the approach.
We start with a reference implementation of the solver in Flash-X and use an LLM to translate each mathematically relevant step of the algorithm into a corresponding mathematical notation.  Note that these are not the equations that one would encounter in an algorithm paper. They detail all the computations that must take place in the algorithm down to the smallest detail in a step-by-step exposition of the arithmetic in the solver. We need this degree of detail to fully verify the new implementation against the reference implementation. A direct implementation from this derivation can be verified against the reference implementation, and any mismatch in the derivations can be corrected in the specification. 

Starting from this description it becomes possible to identify reusable patterns in the solver and encode them into a condensed vocabulary of symbolic representations specific to the solver. One needs to create tools to lower the symbolized representation into a concrete form in the target language for compilation, and this is where the availability of highly capable LLMs changes the cost-benefit equation. 
As mentioned earlier, the existing portability solutions were constrained by the limitation of resources in the form of human developers to even entertain the idea of specialized symbolization most suitable for a single solver or application. They had to necessarily design their solutions to be general across a variety of solvers.
The LLMs have largely eliminated this trade-off to the point where tools for lowering abstractions to a compilable language
can be generated and verified even by developers with no compiler expertise. 

\begin{figure}
    \centering
    \includegraphics[width=0.85\textwidth]{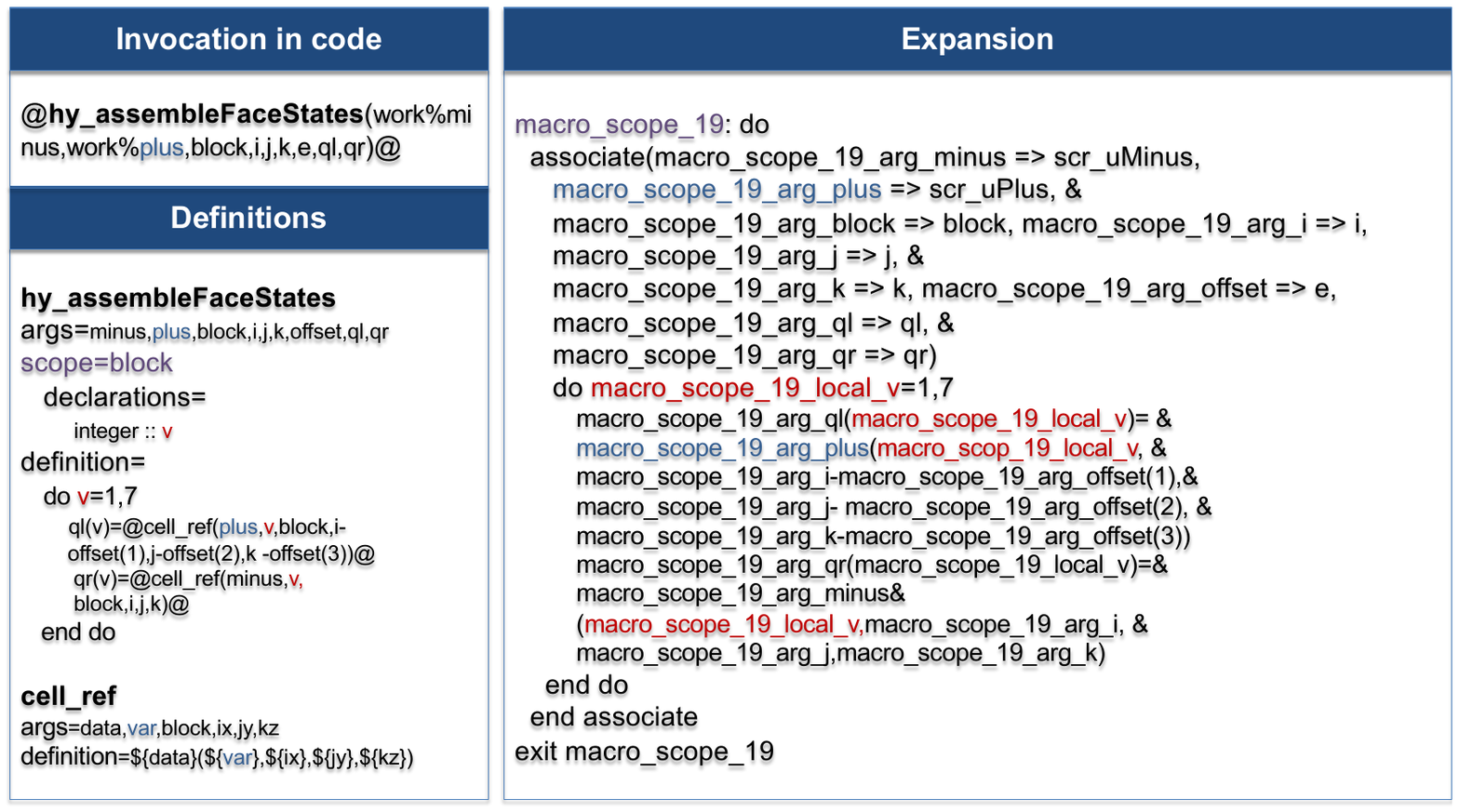}
     \captionsetup{skip=-20pt}
    \caption{An Example of a computational helper. The top left box shows how it is invoked in the code, the bottom left box shows the definitions of the macros used and the right box shows the expanded code.}
    \label{fig:macros}
    \vspace{-0.15in}
\end{figure}

As with any concrete implementation of a numerical algorithm, there is no unique way to design the vocabulary and there is no binding to any specific language when equations are the starting
point. 
We opted to use macros for symbolizing data references as with the first example described in Section~\ref{sec:dataref-only}, but we extended their role and also turned them into computational helpers. These computational helpers are templates that encode the algorithmic sequence of the solver with insertion markers for imports and declarations in the enclosing specification section.
Helper definitions have a name, formal arguments, declarations when needed, and an executable body. Each of the helper also implies a scope that effectively becomes a unique namespace for the translator. Figure \ref{fig:macros} shows an example of a helper that assembles face states. Selected color coded symbols in the figure show how the symbols are placed in the maintained abstracted code, how they are defined, and what do they look like when translated. Note that the translated symbols look very similar to how the linkers resolve symbols. The translator essentially performs a similar function, although at a much more limited scale.  

The expander hoists generated declarations to these markers, placing imports where appropriate in the generated code.
Scratch storage is caller-owned and reusable across calls. Layout-specific allocation is performed by the caller through backend macros. Mesh services are not called directly: the whole-step (including all stages in a multi-stage method) driver receives a provider context and the procedures that are needed to perform these services. 

%% file: experiments.tex
\providecommand{\NI}{\textsf{NewImpl}}
\providecommand{\UM}{\textsf{UniformMesh}}
\providecommand{\CM}{\textsf{CMacro}}

\section{Verification}
\label{sec:verification}
We evaluate the abstracted code from two perspectives: verification -- whether the arithmetic layer produces the same numerical results regardless of who owns the data and drives the time integration; and performance -- whether the generated surrounding code can be tailored for a target platform to obtain good  performance. 

\subsection{Verification of symbolic data-reference}
To verify the abstractions where only the explicit data-references are symbolized we use only the Sedov problem. Because the abstractions are shallow, verification with any one problem suffices. We use two alternative sets of definitions of macros that change the order of indices in the state data layout. The first set of definitions assumes the data layout to be $U(field\_variable,i,j,k)$ where $i,j,k$ are space indices and $field\_variable$ represents physical variables (i.e. density, pressure etc). For the same set of indices the second set of definitions assumes the layout to be $U(i,j,k,field\_variable)$. We refer to these definitions as variable-first and variable-last from here on. Note that Flash-X already has a built-in capability to change the order of indices during configuration because it supports two mesh packages, Paramesh \cite{macneice2000paramesh}, which assumes variable-first order, and AMReX \cite{AMReX_JOSS}, which  assumes variable-last order. For the purpose of verifying the symbolic data referencing we disabled the native reordering capability in Spark and used the LLM to replace all explicit array references in the solver with appropriate symbolic representations. 

\label{sec:exp1}
\begin{wrapfigure}{r}{0.46\linewidth}
    \centering
    \vspace{-10pt}
    \includegraphics[
        width=\linewidth,
        trim=300 100 20 50,
        clip
    ]{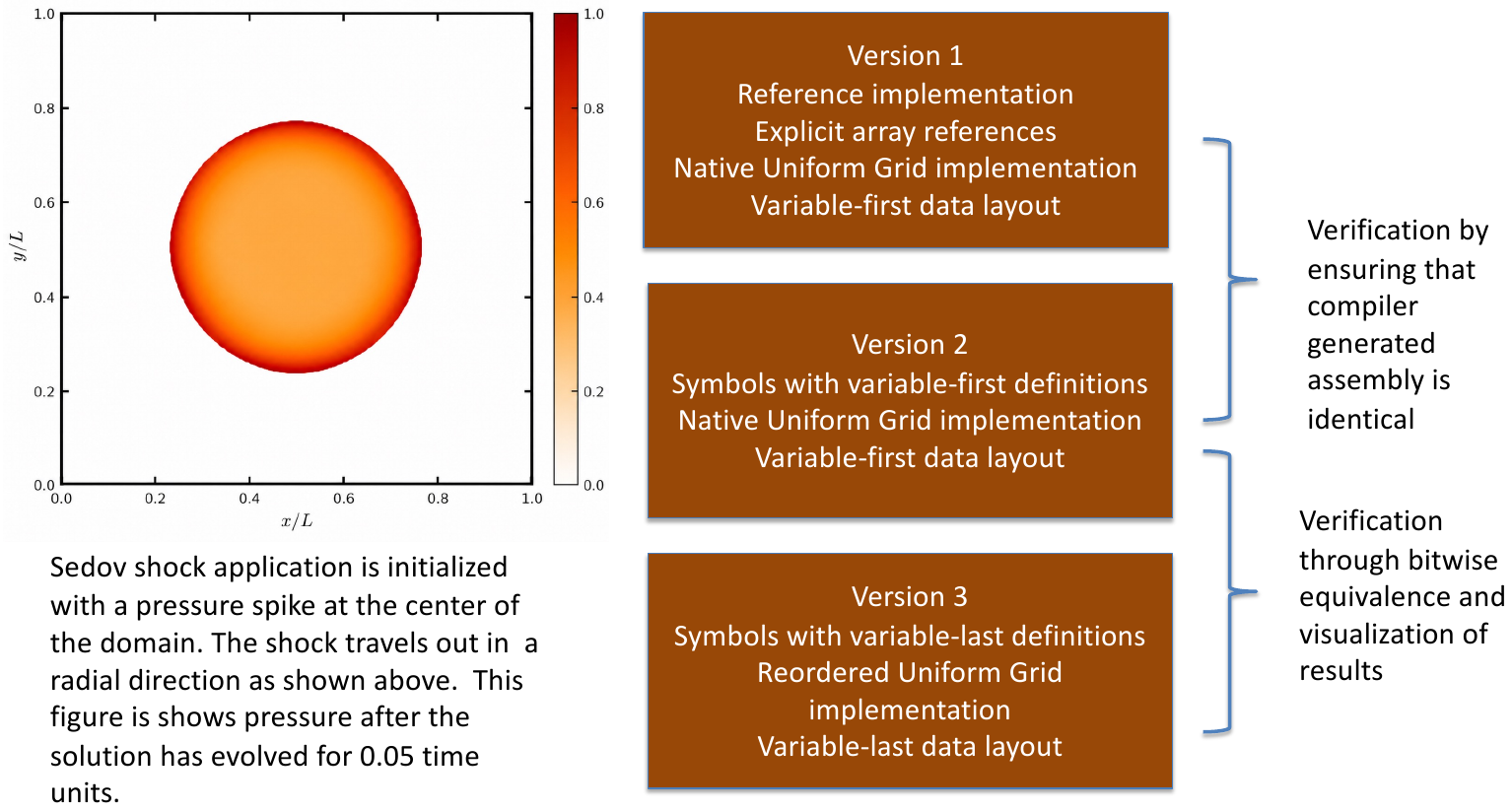}
    \caption{Sedov blast problem run with three different configurations: the reference, abstracted with layout identical to the reference, and abstracted with a different data layout}
    \label{fig:schematic}
    \vspace{-10pt}
\end{wrapfigure}

This set of verification runs used a Mac M5 laptop with Tahoe 26.5 operating system.  The existing mechanism for selection of alternative implementation in Flash-X was used to generate three versions of the problem shown in the figure \ref{fig:schematic}. The first two versions have identical data layout. Their equivalence was verified by confirming that the assembly code produced by the compiler was identical for all the relevant files. The equivalence of the second and the third versions was verified by confirming that the produced results were bit-wise identical, and also through visualization of the results.

\subsection{Verification for Arithmetic Extraction}
This set of experiments probe the efficacy of separating the core arithmetic of the solver from the surrounding code. We refer to the version of the code with its core arithmetic extracted as \textsf{NewImpl} for convenience. The questions we ask are:
\begin{enumerate}
    \item[Q1] Does \textsf{NewImpl} reproduce the results of the reference implementation when it is inserted into Flash-X?
    \item[Q2] Does \textsf{NewImpl} produce the same results when its data are owned and its time loop is driven by an independently developed infrastructure instead of Flash-X?
    \item[Q3] How much code has to be written to attach an independent infrastructure, and does any of it touch the arithmetic layer?
    \item[Q4] Does the arithmetic layer give the same results on a GPU?
\end{enumerate}

\begin{figure}[htbp]
    \centering
    \includegraphics[width=0.75\textwidth]{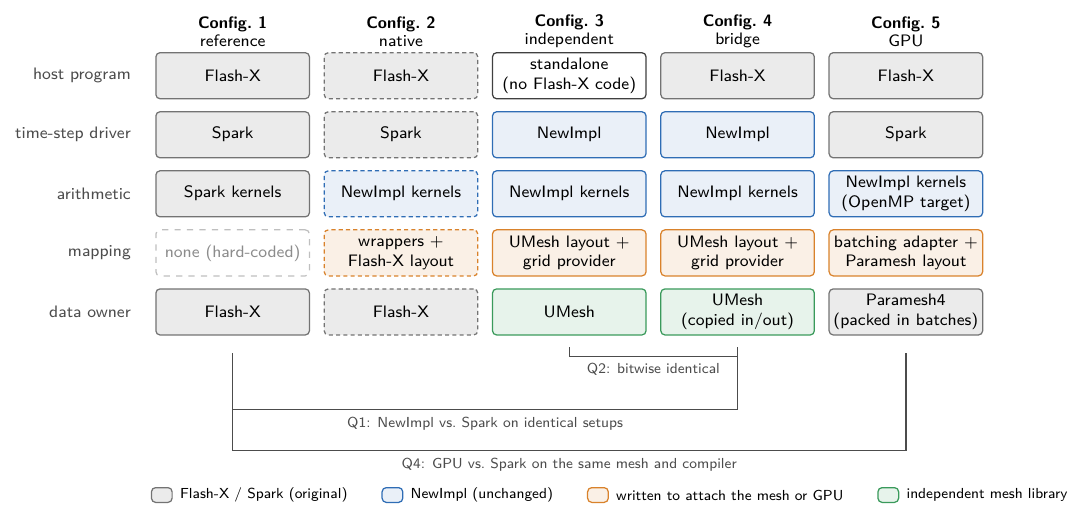}
    \caption{Evaluated configurations. "Data owner" is the software that allocates and lays out the cell data; "time-step driver" is the code that sequences Runge--Kutta stages, guard-cell exchange, and flux synchronisation within a time step.}
    \label{fig:configurations}
\end{figure}

We compare five configurations described in Figure~\ref{fig:configurations}. They differ in who owns the cell data (and thus also the layout in memory), who drives the time step, and which hydrodynamics implementation performs the arithmetic. Here, the reference configuration, {\em config 1}, is the unmodified Spark solver running with Flash-X's native UG. The second configuration, {\em config 2}, embeds the core arithmetic from \textsf{NewImpl} in Spark's own driver through compatibility wrappers that we developed with LLM assistance. 
We run this configuration on Flash-X's single UG block, like {\em config 1}.

The third configuration, {\em config 3}, runs with core arithmetic of \textsf{NewImpl} with independent infrastructure.  Here, we use a mesh developed as a C library with Fortran binding that divides a uniform Cartesian domain into equal blocks with halos, and whose memory layout, dimension, and halo width are compile-time macros. \textsf{NewImpl}'s whole-step driver advances the solution and \textsf{UniformMesh} supplies the data and the
mesh services through a grid provider. The host program contains no Flash-X code. It reproduces Flash-X's initial conditions, 
time-step control, and output formats from Flash-X's sources, so that its results can be compared with {\em config 1} and {\em config 4} using Flash-X's own tools.

In the fourth configuration, {\em config 4}, we embed the whole of \textsf{NewImpl} and the independently developed mesh as a sub-unit of Spark in Flash-X. Here, Flash-X initializes the problem and controls the time step; in each step its cell data are copied into \UM, advanced by \NI, and copied back. The configuration allows \textsf{NewImpl} and Spark to be verified with bit-wise identical results. The final configuration, {\em config 5}, uses OpenMP target offload to enable running the code on the GPU. In this configuration, NewImpl's core arithmetic is embedded into the single stage face-flux calculation routine of Spark. Unlike all previous configurations, here the single stage driver for face-flux calculation needs to be modified to allow batching of blocks for offloading. This is because in AMR mode an individual block is too small for efficient use of the GPU. Several blocks must be computed in parallel on the GPU for   it to be viable in production.  To verify this version, the same generated source is also built for host execution (OpenMP on the CPU, "host kernels"). The mesh is a single refinement level of $4\times4$ blocks of 64\textsuperscript{2} cells, so it has the same 256\textsuperscript{2} resolution as the other configurations, with the periodic boundaries and flux correction. We verify this configuration against the reference configuration using with the same compiler on the same mesh.

For all of these verification runs \textsf{NewImpl} and Spark are taken from the same commit of Flash-X; \textsf{NewImpl}'s sources are not modified in any configuration, which each build verifies by recording checksums of all \textsf{NewImpl} inputs. All codes that have CPU only target are compiled with GNU Fortran~14.3.1 and Flash-X's flags (\texttt{-O2}, 8-byte reals) and run on a single core of an AMD EPYC~7302 node, one MPI rank. 
{\em Config 5} and its Spark reference are compiled with NVHPC 26.3 (-O2 -r8, -mp=gpu -gpu=cc128) and run on an NVIDIA GeForce RTX5080 attached to an Intel Xeon Max 9468, with one MPI rank. Performance measurements on the GPU were conducted on a system with NVIDIA GB200 NVL4 with NVHPC 26.3 compiler suite.

\subsubsection{Details of the runs}
We use Sod shock tube and isentropic vortex problems for the CPU only runs. All variations of these problems run in two-dimensions with periodic boundaries on $256\times256$ cells. \texttt{IsentropicVortex} was setup on the physical domain sized at $[0,10]^2$, advected diagonally for one period ($t=10$), where the exact solution is the translated initial condition. 
The \texttt{Sod} setup used a physical domain of size $[0,1]^2$ with the interface normal to $x$, run to $t=0.025$. With periodic boundaries a second, mirror-image Riemann problem forms at $x=0$, so the exact solution is the superposition of two exact Riemann solutions until the wave systems meet, and it is symmetric about $x=0.25$. 
A second variant (\emph{Sod-hybrid}) switches to the HLLE solver in shock-flagged cells. Two further variants switch flattening off in both solvers (\emph{Sod-noflat}, \emph{Sod-hybrid-noflat}), to identify differences at shocks. For the order of accuracy, the vortex is also run on $64^2$, $128^2$, and $512^2$ cells with the remaining parameters left identical to the other runs. 
$1.84\times10^{-5}$ / $4.0\times10^{-4}$ 
Two vortex variants use a fixed $\Delta t=5\times10^{-3}$ after Flash-X's start-up step ($0.1$ of the CFL limit). One runs a full period (\emph{vortex-fixed}, 2001 steps), the other 100 steps (\emph{vortex-short}).

We run {\em config 3} and {\em config 4} with several decompositions of the domain into blocks of at least $8\times8$ cells, from strips ($4\times1$, $1\times16$) to $32\times32$ blocks. Each problem uses two to six of them and each run is compared to the single block run of the same problem. 20 comparisons per configuration, 16 of them on the five main problems. The time per step is measured on vortex-short (100 steps) and Sod (30 steps), five times each on an otherwise idle node; we report the minimum.


To determine which of the Flash-X behaviours reproduced by {\em config 3} bit-wise identity depends on, configuration~3 is rerun with one of them replaced at a time and compared with configuration~4 on vortex-short, Sod, and the vortex: Flash-X's equation of state by the grid provider's simpler one, or without recomputing the internal energy; no start-up step; no limit
on the growth of the time step; the last step shortened to the final time; and \NI's own CFL formula instead of Spark's. To exercise storage layouts \NI's own example, a one-dimensional periodic Sod problem on its built-in grid provider, is built with each of \NI's three storage mappings (variable-first, spatial-first, and per-cell records) and run with SSP-RK2 and SSP-RK3 on one and four blocks; the outputs are compared byte by byte.


\subsubsection{Metrics}
\begin{itemize}
  \item \emph{bit-wise identity} of checkpoints, compared with Flash-X's
    \texttt{sfocu} tool at tolerance zero, and of the time series of integral quantities (which include the simulation time and thus the time-step sequence).
  \item \emph{Difference between two configurations}: the maximum and $L_1$
    norm of the cell-wise difference of all variables (from \texttt{sfocu}),
    and its ratio to the discretisation error.
  \item \emph{Error against the exact solution}: $L_1$ and $L_\infty$ norms of the density error, with the exact solution evaluated at each checkpoint's own time.
  \item \emph{Mirror symmetry} of the Sod solution:
    $\max_x |u(x)+u(x')|$, where $x'$ is the mirror image of $x$ about
    $x=0.25$; it is zero for the exact solution.
  \item \emph{Number of time steps} and the time-step sequence.
  \item \emph{Lines of code} that attach the infrastructure, and lines of the arithmetic layer changed for it.
  \item \emph{Conservation}: the largest relative change of the total mass, momentum, and energy over a run.
  \item \emph{Time per step} of the hydrodynamics: Flash-X's \texttt{Hydro} timer for configurations~1 and~4 (for configuration~4 including the copies into and out of \UM), and \NI's whole-step driver for configuration~3; and the wall time of the whole run per step.

\end{itemize}


%% file: results.tex
\subsection{Verification Results}
\label{sec:results}

\begin{table}[t]
  \caption{\NI{} on \UM{} in Flash-X (config~4) versus the original Spark solver (config~1); $256\times256$ cells, one rank. Differences are maxima over all cells and variables; errors are density errors against the exact solution ($L_1$ / $L_\infty$). 
  vortex-short: $L_1$ difference $2.5\times10^{-15}$.}
  \label{tab:ni-spark}
  \centering\small\setlength{\tabcolsep}{4pt}
  \begin{tabular}{@{}lrlll@{}}
    \toprule
    Problem & Steps & $|\text{\NI}-\text{Spark}|_{\max}$ & \NI{} error & Spark error \\
    \midrule
    vortex-short & 100 & $1.9\times10^{-12}$ & $2.25\times10^{-6}$ / $8.9\times10^{-5}$ & same \\
    vortex-fixed & 2001 & $1.3\times10^{-8}$ & $1.84\times10^{-5}$ / $4.0\times10^{-4}$ & same \\
    vortex (CFL) & 1877 & $2.5\times10^{-8}$ & $1.84\times10^{-5}$ / $4.0\times10^{-4}$ & same \\
    Sod & 30 & $1.6\times10^{-13}$ & $3.78\times10^{-3}$ / $8.47\times10^{-2}$ & same \\ 
    Sod-hybrid & 30 & $1.6\times10^{-13}$ & $3.79\times10^{-3}$ / $8.54\times10^{-2}$ & same \\ 
    \bottomrule
  \end{tabular}

\end{table}

We begin by comparing the reference solver with core arithmetic of \NI{} ({\em config 1} and{\em config 4}), summarized in Table~\ref{tab:ni-spark}. For smooth flow after 100 identical steps the two solvers differ by at most $1.9\times10^{-12}$ in single cells, with an $L_1$ difference of $2.5\times10^{-15}$, which can be considered as rounding errors. 
Over a full period the maximum difference grows to $1.3\times10^{-8}$ with a fixed time step and $2.5\times10^{-8}$ with the CFL-limited timestep. That is $0.003\%$ and $0.006\%$ of the maximum error against the exact solution. Even with the CFL-limited timestep, both solvers choose identical time-step series and have the same error against the exact solution.
For smooth flow \NI{} reproduces Spark, including its time-step control, to within rounding effects (Q1).

On the Sod problem, \NI{} reproduces Spark to rounding. The maximum difference is $1.6\times10^{-13}$ (with and without the hybrid HLLE solver). Both solvers take the same 30 time steps, have the same error against the exact solution and preserve the mirror symmetry of the periodic problem to $\max|u(x)+u(x')| = 10^{-13}$.
Shock detection, both Riemann solves, flattening, artificial viscosity and the update thus agree with Spark's implementation.

\subsubsection{Independent infrastructure (config 3 versus 4)}
\label{sec:res-standalone}
Configuration~3 runs \NI{} on \UM{} without any Flash-X code. On the same machine and compiler, its results are bit-wise identical to those of configuration~4 for all ten problems: 
every checkpoint (11 for each vortex run over a full period, compared with \texttt{sfocu} at tolerance zero), the time series of integral quantities including the simulation time, and therefore the entire time-step sequence. This answers Q2 -- the result of the arithmetic does not depend on the host program. Mimicking  Flash-X's initialization, time-step control and output by an independent implementation produces bit-wise identical results. 


\begin{table}[t]
  \caption{Ablation of configuration~3: maximum difference to configuration~4 in the hydrodynamic variables when one reproduced Flash-X behaviour is replaced; -- means bit-wise identical.}
  \label{tab:ablation}
  \centering\small\setlength{\tabcolsep}{4pt}
  \begin{tabular}{@{}llll@{}}
    \toprule
    Replaced behaviour & vortex-short & Sod & vortex \\
    \midrule
    Equation of state of the grid provider & $3.0\times10^{-12}$ & $1.2\times10^{-13}$ & $2.7\times10^{-8}$ \\
    No recomputation of the internal energy & $4.4\times10^{-13}$ & $3.6\times10^{-14}$ & $2.3\times10^{-8}$ \\
    No start-up step & $3.3\times10^{-3}$ & -- & $7.4\times10^{-4}$ \\
    No limit on time-step growth & -- & $1.1\times10^{-2}$ & $1.6\times10^{-3}$ \\
    Last step shortened to final time & -- & $1.0\times10^{-1}$ & $3.8\times10^{-3}$ \\
    \NI's CFL formula & -- & -- & $2.8\times10^{-8}$ \\
    \bottomrule
  \end{tabular}
\end{table}

\subsubsection{Native environment (Config 2 vs config 1).}
\label{sec:res-native}
Inside Spark's own driver, \NI{}'s kernels reproduce Spark to rounding on every problem. After a full vortex period with a fixed time step the maximum difference is $1.7\times10^{-12}$, on the Sod problems $3.0\times10^{-13}$ ($2.4\times10^{-13}$ with the hybrid solver), with identical step counts.
The only larger difference, $3.0\times10^{-8}$ for the CFL-limited vortex, arises because of rounding form the first steps on, so the time steps computed from them differ in the last bits and this error grows to the same order as for configuration~4 ($2.5\times10^{-8}$). With a fixed time step the difference stays at $1.7\times10^{-12}$
The arithmetic layer thus behaves identically in its native environment and in the independently developed infrastructure.

\subsubsection{Different architecture (Config 5 vs config 1).}
\label{sec:res-architecture}
Table~\ref{tab:gpu-spark} compares the GPU configuration with Spark built with the same compiler on the same mesh. On every problem both take the same number of steps. With a fixed time step they take the identical time-step sequence. On Sod the maximum difference is $3.4\times 10^{-13}$. Mirror symmetry is preserved to $3\times10^{-13}$, the same level as Spark itself. After 100 vortex steps the difference is $5.2\times^{-12}$. 
To separate the kernels from the hardware, the generated kernels were also run on the host. With contraction disabled, the GPU results are bit-wise identical to the host results on all five problems. The results are also independent of the batch size (1, 3 and 32).

\begin{table}[t]
  \caption{\NI{} on a GPU (configuration~5) versus the original Spark solver, both built with NVHPC~26.3 on the same Paramesh4 mesh ($4\times4$ blocks of $64^2$ cells, one level); one rank. Differences are maxima over all cells and variables of the final checkpoint; $L_1$ is the mean cell-wise difference. ''Contraction off'': both codes built without fused multiply-add contraction. Last column: the GPU run compared with the same generated kernels executed on the host, both without contraction.}
  \label{tab:gpu-spark}
  \centering\small\setlength{\tabcolsep}{4pt}
  \begin{tabular}{@{}lrllll@{}}
    \toprule
    Problem & Steps & $|\text{GPU}-\text{Spark}|_{\max}$ & $L_1$ & contraction off & GPU vs.\ host \\
    \midrule
    vortex-short & 100  & $5.2\times10^{-12}$ & $4.0\times10^{-15}$ & $3.8\times10^{-12}$ & bit-wise \\
    vortex-fixed & 2001 & $1.4\times10^{-7}$  & $2.3\times10^{-9}$  & $1.9\times10^{-7}$  & bit-wise \\
    vortex (CFL) & 1877$^{a}$ & $6.9\times10^{-8}$  & $1.0\times10^{-9}$  & $8.2\times10^{-8}$  & bit-wise \\
    Sod          & 30$^{a}$   & $3.4\times10^{-13}$ & $7.1\times10^{-15}$ & $2.4\times10^{-13}$ & bit-wise \\
    Sod-hybrid   & 30$^{a}$   & $2.3\times10^{-13}$ & $5.8\times10^{-15}$ & $9.3\times10^{-14}$ & bit-wise \\
    \bottomrule
  \end{tabular}

  \smallskip
  \raggedright\footnotesize $^{a}$Same number of steps as Spark; the CFL time
  steps differ only in the last digits (end times by at most
  $2.5\times10^{-12}$ for the vortex and $10^{-17}$ for Sod).
\end{table}

\subsection{Effort to attach the infrastructure}
\label{sec:res-effort}
Table~\ref{tab:effort} lists the code written to attach \UM{} to \NI{} (Q3).
The arithmetic layer, 1102 lines of \NI{} templates, helper definitions, and the grid-provider contract, is used without any changes.
Two files connect \NI{} to \UM{}. The storage mapping (61 lines) defines the field type and the \texttt{cell\_ref} and the \texttt{face\_ref} references as a pointer into \UM{}'s buffer, so no cell data is copied. Scratch fields remain ordinary arrays. The grid provider (194 lines) implements the four services of the contract: guard-cell exchange through \UM{}, flux synchronisation at block borders, the global reductions (identity for one rank), and the equation of state. 
Everything else in Table~\ref{tab:effort} belongs to a host program, not to the attachment (the Flash-X sub-unit of configuration~4 and the
standalone program of configuration~3, most of which reproduces Flash-X's initialisation, time-step control, and output formats).
Running \NI{} on a GPU required 732 lines of new code (Table~\ref{tab:effort}), none of which touches the arithmetic layer.

\begin{table}[t]
  \caption{Lines of code (including comments) per component. Only the first
    two rows are needed to attach \UM{} to \NI; only the rows marked 5 are
    needed to run \NI{} on a GPU. The GPU kernels themselves (841 lines) are
    generated from the unchanged \NI{} templates and are not maintained.}
  \label{tab:effort}
  \centering\small
  \begin{tabular}{@{}lrl@{}}
    \toprule
    Component & Lines & Used by \\
    \midrule
    Storage mapping (layout definitions) & 61 & 3, 4 \\
    Grid provider (contract implementation) & 194 & 3, 4 \\
    \midrule
    GPU kernel template and device layout definitions & 140 & 5 \\
    GPU kernel generator & 200 & 5 \\
    Batching adapter (packing, transfers, launches, flux publication) & 392 & 5 \\
    \midrule
    Flash-X sub-unit and installer & 588 & 4 \\
    Standalone host program (incl.\ checkpoint writer) & 1767 & 3 \\
    \UM{} library (C, Fortran binding, layout macros) & 884 & 3, 4 \\
    \midrule
    \NI{} arithmetic layer, lines changed & 0 of 1102 & 3, 4, 5 \\
    \bottomrule
  \end{tabular}
\end{table}

\section{Performance}
We analyzed the performance of the code from two perspectives, that of overheads and portability. 
Table~\ref{tab:timing} lists the time per step. Expressing the arithmetic through application-defined abstractions has no measurable cost. Within Flash-X, \NI{} takes $0.127$\,s per step on the vortex where Spark takes $0.140$\,s. Configuration~3 takes $0.111$\,s. The difference to configuration~4 is not in \NI: configuration~4's timer additionally covers copying the solution between Flash-X's array and \UM{} in every step, which the bridge configuration needs because both own a copy of the data, and Flash-X's general equation-of-state interface, which gathers the cell data into separate vectors; configuration~3 evaluates the same equation of state in place.
Outside the hydrodynamics, Flash-X adds 0.02--0.06\,s per step for time-step control and output, the standalone host almost nothing. These are single-core measurements on one machine with different timer scopes. 

\begin{figure}[htbp]
    \centering
    \includegraphics[width=0.7\textwidth]{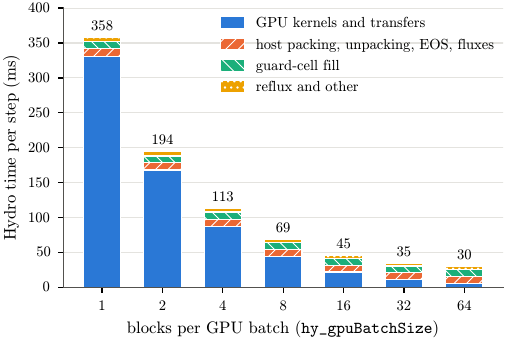}
    \caption{Hydrodynamics time per step on the GPU for Sedov (256 blocks of $16^2$ cells, 100 steps) versus the number of blocks per GPU batch, split into GPU kernels and transfers, host-side adapter work, guard-cell exchange, and the rest.}
    \label{fig:batching-breakdown}
\end{figure}

\begin{figure*}
  \includegraphics[width=\textwidth]{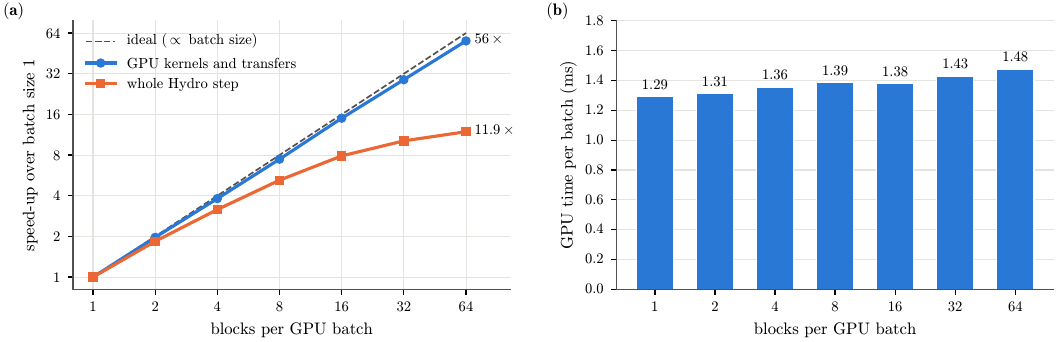}
  \caption{(a) Speed-up over batch size 1 of the GPU kernels and transfers and of the whole hydrodynamics step. (b) GPU time per batch, almost independent of the number of blocks in it.}
  \label{fig:batching-speedup}
\end{figure*}

 On a GPU the surrounding code decides how much work each kernel launch gets. Figure~\ref{fig:batching-breakdown} shows the time per step on Sedov for batch sizes 1 to 64. A batch costs 1.29 ms with 1 block and 1.48 ms with 64 (Figure~\ref{fig:batching-speedup}), so each launch is dominated by launch overhead rather than computation. Grouping blocks therefore cuts the GPU time almost in proportion to the batch size. 
The time for the whole hydrodynamics step falls from 358 ms to 30ms ($11.9\times$). At that point it is limited by host work that batching does not reduce. About 10 ms for guard-cell exchange and about 9 ms for packing, unpacking, the equation of state and fluxes. The arithmetic layer is the same for every batch size and gives bit-wise identical results. These measurements clearly show that the surrounding code can be adapted to an architecture's execution model.

\begin{table}[t]
  \caption{Time per step in seconds (minimum of five runs), $256\times256$ cells, one core. ``Hydro'': Flash-X's \texttt{Hydro} timer configurations 1, 4) or \NI{}'s whole-step driver (configuration 3); ``wall'': whole run per step.}
  \label{tab:timing}
  \centering\small\setlength{\tabcolsep}{4pt}
  \begin{tabular}{@{}lllll@{}}
    \toprule
    & \multicolumn{2}{l}{vortex-short} & \multicolumn{2}{l}{Sod} \\
    Configuration & Hydro & wall & Hydro & wall \\
    \midrule
    1: Spark & 0.140 & 0.159 & 0.140 & 0.198 \\
    4: \NI{} in Flash-X, 1 block & 0.127 & 0.146 & 0.124 & 0.183 \\
    4: \NI{} in Flash-X, $4\times4$ / $4\times1$ & 0.131 & 0.150 & 0.136 & 0.195 \\
    3: \NI{} standalone, 1 block & 0.111 & 0.114 & 0.109 & 0.113 \\
    3: \NI{} standalone, $4\times4$ / $4\times1$ & 0.117 & 0.121 & 0.112 & 0.117 \\
    \bottomrule
  \end{tabular}
\end{table}

